\documentclass{IOS-Book-Article}

\usepackage{mathptmx}
\usepackage{soul}\setuldepth{article}
\usepackage{graphicx}

\usepackage{hyperref}
\hypersetup{colorlinks=true, linkcolor=black, urlcolor=black}

\usepackage{algorithm}
\usepackage{algpseudocode}
\usepackage{amsmath}

\usepackage{xcolor}

\newcommand{\mycomment}[1]{}

\def\hb{\hbox to 11.5 cm{}}

\begin{document}

\pagestyle{headings}
\def\thepage{}

\begin{frontmatter}              

\title{Network Density Analysis of Health Seeking Behavior in Metro Manila: \\ A Retrospective Analysis on \\ COVID-19 Google Trends Data}


\author[A]{\fnms{Michael T.} \snm{LOPEZ II}
\thanks{Corresponding Author: Michael T. Lopez II, Department of Information Systems and Computer Science, Ateneo de Manila University, Quezon City, Philippines; E-mail: michael.lopez@student.ateneo.edu}}
\author[A]{\fnms{Cheska Elise} \snm{HUNG}}
and
\author[A]{\fnms{Maria Regina Justina E.} \snm{ESTUAR}}

\address[A]{Department of Information Systems and Computer Science, \\ Ateneo de Manila University, Philippines}

\begin{abstract}
This study examined the temporal aspect of COVID-19-related health-seeking behavior in Metro Manila, National Capital Region, Philippines through a network density analysis of Google Trends data. A total of 15 keywords across five categories (English symptoms, Filipino symptoms, face wearing, quarantine, and new normal) were examined using both 15-day and 30-day rolling windows from March 2020 to March 2021. The methodology involved constructing network graphs using distance correlation coefficients at varying thresholds (0.4, 0.5, 0.6, and 0.8) and analyzing the time-series data of network density and clustering coefficients. Results revealed three key findings: (1) an inverse relationship between the threshold values and network metrics, indicating that higher thresholds provide more meaningful keyword relationships; (2) exceptionally high network connectivity during the initial pandemic months followed by gradual decline; and (3) distinct patterns in keyword relationships, transitioning from policy-focused searches to more symptom-specific queries as the pandemic temporally progressed. The 30-day window analysis showed more stable, but less search activities compared to the 15-day windows, suggesting stronger correlations in immediate search behaviors. These insights are helpful for health communication because it emphasizes the need of a strategic and conscientious information dissemination from the government or the private sector based on the networked search behavior (e.g. prioritizing to inform select symptoms rather than an overview of what the coronavirus is).
\end{abstract}

\end{frontmatter}

\section{Introduction}

The emergence of the COVID-19 pandemic in the Philippines was a landmark point in the public health landscape of the nation. The rapid spread of the virus, characterized by its high rate of contagion and the challenges posed by overlapping symptoms with other respiratory illnesses, required urgent and comprehensive responses from health authorities \cite{amitEarly2021}. In an effort to lessen the pandemic's impact, the Philippine government implemented new community lockdowns \cite{bbcnewsCoronavirus2020}. The national economic toll was as severe, with incurred financial losses at an estimated amount of 2.9 to 3.2 trillion Philippine Pesos (PhP) between 2020 and 2021 \cite{delcastilloEstimating2024}. Thus, the public sought information on online platforms due to the surrounding uncertainty. For example, Filipinos exhibited a high propensity to believe in misinformation, particularly about the alternative treatment of Ivermectin, which witnessed a surge in Google searches during spikes in COVID-19 cases \cite{alibudbudCase2023}. Interestingly, the pandemic caused changes in consumer behavior, with increased impulse buying as coping mechanisms \cite{mallariMediating2023}. The significance of understanding public sentiment became more evident. Since then, researchers further investigated how people would react to the coronavirus pandemic from innovative data sources such as Google Trends \cite{alibudbudGoogle2023}.

Given the vast amount of resources on the Internet due to its worldwide exponential growth, there is potential in how ``big data'' could be fully analyzed in health informatics \cite{alibudbudGoogle2023}. Google Trends is a free and open-source platform that provides relative popularity depending on the time and location attributes set by the user \cite{rovettaGoogle2024}. The numerical value of the query’s popularity is called the \emph{relative search index} (RSV) \cite{cebrianAddressing2024}. This data point is widely used in infodemiological studies because it provides real-time datasets about user behavior made available to anyone \cite{lolicDIY2024}. It is the leading source in applied health research to supplement decision-making for government officials because it could support predicting future disease cases \cite{rovettaGoogle2024}. \mycomment{Such examples of these illnesses were dengue \cite{syamsuddinCausality2020}, hand-foot-mouth disease \cite{santangeloDigital2023}, the seasonal flu \cite{rahulUnveiling2023}, monkeypox \cite{jokarMonkeypox2023}, and pancreatic cancer among others \cite{gianfrediUsing2023}.} There is also evidence that the prediction model will most likely be improved with the aid of Google Trends variables. In short, it outperforms traditional time series models \cite{behnenExperimental2020}. These points are evidences that augmenting online search engine data with epidemiology is a trusted method to implement. One of the reasons why is rooted to the rising number of users that actively track their life online. Some even depend on their daily routines with it \cite{lolicDIY2024}.

However, there are limitations to how Google Trends returns data when requested. This is in connection with the proprietary nature of the platform \cite{mavraganiGoogle2019}. The most highlighted weakness is how the popularity index scores are based on a subset of the entire search population at a given time \cite{cebrianAddressing2024}. This sampling variability could be detrimental to the results of an experiment based on the collected data because it is caused by a \emph{random draw} \cite{eichenauerObtaining2022}. Taking into account how Google Trends suffers from inconsistent data returns, there is a chance that when someone is asking for the same database attributes (similar region, period, and keyword), but applied on a different date and time, it could generate a whole new set of values \cite{rovettaGoogle2024}. One study proved that the term ``GDP growth'' when it was returned from Google Trends with geographic country filters of ``Brazil'' and the ``United States'' had fluctuating RSV values when collected on different dates and times. It was shown that the three distinct RSV time-series in Brazil had correlations between 0.49 and 0.56 against each other. Then, correlation values of between 0.51 and 0.65 for the data from the United States \cite{medeirosProper2021}. Given the obstacle, this study tests a novel means on how to mitigate the anomalies in the RSV values \cite{chuEnhancing2023}. Instead of focusing on one search query, it is proposed to consider social network statistics on a group of keywords, such as network density and clustering coefficient \cite{heveyNetwork2018, bedruBig2020}. This social network analysis strategy was also applied for contact tracing amidst the pandemic \cite{soTopological2021}. The purpose of viewing them as a network is to determine what different keywords are related to each other in terms of the search behavior at that timespan. \mycomment{It shifts the importance away from the RSV values into the \emph{correlation} scores of one keyword to another.}Thus, instead of focusing on an individual keyword at a time, this method will simultaneously consider other related queries that concern the coronavirus pandemic. 

Another challenge is how could the user overcome the \emph{limited temporal intervals} of a returned Google Trends data. For instance, it is impossible to request for a year-long daily time-series data of a Google Trends search term. This is because there is a limit of until nine months to view its corresponding daily data. If the user requested a time period beyond nine months, the data returned is now available on weekly intervals. Then, Google Trends will give the data on a monthly basis if the temporal setting is beyond 5.25 years \cite{eichenauerObtaining2022}. There are irregularities about how the data is presented in terms of its hourly, daily, weekly, and monthly formats \cite{rovettaGoogle2024}. The time period is one of the filters available to the user for retrieving the data, and it is consequential for its granularity \cite{cebrianAddressing2024}. Therefore, there are repercussions for long-term RSV trends because the resolution of the data became diluted, and it may not be reliable to begin with \cite{behnenExperimental2020, eichenauerObtaining2022}. 

\mycomment{The paper now aims to probe the following research questions:
\begin{enumerate}
    \item How to interpret the computation of the network density and clustering coefficient of all keywords in relation to the general search behavior of COVID-19 keywords in Metro Manila?
    \item How does the threshold value affect the respective computed network density and clustering coefficient time-series?
    \item Why it is important there are proposals to circumvent the limitations of Google Trends in terms of downloading the query's dataset?
\end{enumerate}}



\section{Methodology}

This section features the procedure of how the keywords were transformed into a network graph. \mycomment{It is divided into three parts. To aid in understanding the different stages, shown in Figure~\ref{fig:framework} is a visual framework of the study's method of fulfilling the objective.} The first step in Section~\ref{sec:first-gtdata} provides a rationale for the words chosen as search queries, and explains the data stitching algorithm on how it became a year-long time-series with daily intervals. Section~\ref{sec:second-corr} shares how the keywords were computed to determine the correlation and adjacency matrix. Finally, Section~\ref{sec:third-networkstat} formally defines the network density and clustering coefficient, and their relevance to the interpretation of the network analysis of the Google Trends time-series data. For clarity, the terms \emph{query}, \emph{search term}, and \emph{keyword} would be used interchangeably in this paper, referring to the word or phrase that the user requested from Google Trends \cite{cebrianAddressing2024}.

\mycomment{\begin{figure}[htbp]
\centering
\includegraphics[width=0.6\textwidth]{method_framework.png}
\caption{Method Framework From Downloading Each Google Trend Query to Its Conversion of Network Graph.}
\label{fig:framework}
\end{figure}}

\subsection{First Stage: Collecting and Preprocess Google Trends Data}\label{sec:first-gtdata}

The study analyzed all 15 keywords into one group. The following were classified into five categories to easily distinguish them. The first category, \textbf{Symptoms (English)}, included the search terms of ``cough,'' ``fever,'' ``flu,'' ``headache,'' and ``rashes.'' The second category, \textbf{Symptoms (Filipino)}, comprised local language terms: ``lagnat'' (fever), ``sipon'' (cold or runny nose), and ``ubo'' (cough). For terms related to preventive measures, the \textbf{Face Wearing} category included searches for ``masks'' and ``face shield,'' while the \textbf{Quarantine} category tracked searches for ``ecq'' (Enhanced Community Quarantine) \cite{departmentofhealthoftherepublicofthephilippinesRecommendations2020} and ``quarantine.'' Finally, the \textbf{New Normal} category monitored adaptation-related terms including ``frontliners,'' ``social distancing,'' and ``work from home.'' Google Trends is not case-sensitive \cite{mavraganiGoogle2019}. So, whether if it is ``ecq'' or ``ECQ'', it will be interpreted similarly. The Google Trends data were retrieved on August 19 to 31, 2024 using the \texttt{pytrends} Python (version 3.12.2) library \cite{Pytrends}. The geographic location was filtered to the ``National Capital Region,'' also known as ``Metro Manila'', since it has the highest percentage of household internet users in the Philippines \cite{philippinestatisticsauthorityMore2023}.

\begin{algorithm}
\caption{Rescaling Annual Google Trends Daily Data with Weekly Data}
\label{alg:rescaling}
\begin{algorithmic}[1]
\Procedure{CalculateWeeklyMetrics}{$W, D$}
   \For{$\text{every week }w \in W$}
       \State $w_s \gets w.\textit{date}$; $w_e \gets \text{next week start}$
       \State $w_d \gets \{d \in D : w_s \leq d.\textit{date} < w_e\}$
       \If{$w_d \neq \emptyset$}
           \State $w.\textit{sum} \gets \sum_{d \in w_d} d.\textit{value}$; $w.\textit{count} \gets |w_d|$
           \State $w.\textit{avg} \gets w.\textit{sum} / w.\textit{count}$
       \EndIf
   \EndFor
\EndProcedure
\Procedure{CalculateWeights}{$W$}
   \For{$\text{every week }w \in W$}
       \State $w.\textit{weight} \gets w.\textit{avg} = 0 ? 1 : w.\textit{value} / w.\textit{avg}$
   \EndFor
\EndProcedure
\Procedure{RescaleValues}{$D, W$}
   \For{$\text{every day }d \in D$}
       \State $w \gets \text{week containing } d.\textit{date}$
       \State $d.\textit{rescaled} \gets w.\textit{avg} = 0 ? d.\textit{value} : d.\textit{value} \times w.\textit{weight}$
   \EndFor
\EndProcedure
\end{algorithmic}
\text{\footnotesize Notes: Expression $a = b ? x : y$ means ``if $a = b$ then $x$, else $y$''. Variables: $W$ -- weekly data, $D$ -- daily data,}
\text{\footnotesize $w_s$ -- week start date, $w_e$ -- week end date, $w_d$ -- daily data in current week, $w.\textit{avg}$ -- weekly average}
\end{algorithm}

The search terms concerning the symptoms were determined using the World Health Organization (WHO) website \cite{worldhealthorganizationAdvice2023, worldhealthorganizationCoronavirus2023}.\mycomment{\footnote{These are the two WHO official sites consulted: \href{https://www.who.int/news-room/fact-sheets/detail/coronavirus-disease-(covid-19)}{www.who.int/news-room/fact-sheets/detail/coronavirus-disease-(covid-19)} and \href{https://www.who.int/emergencies/diseases/novel-coronavirus-2019/advice-for-public}{https://www.who.int/emergencies/diseases/novel-coronavirus-2019/advice-for-public}}} Then, the remaining keywords were decided based on the new established health protocols by the government. Specifically, these are the legally-binding resolutions from the Inter-Agency Task Force for the Management of Emerging Infectious Diseases (IATF-EID),\mycomment{\footnote{IATF Resolutions referenced were from March and April 2020: \href{https://doh.gov.ph/iatf/page/32/}{https://doh.gov.ph/iatf/page/32/}}} which was the executive body that deliberated and enforced the appropriate policies during the COVID-19 crisis \cite{departmentofhealthoftherepublicofthephilippinesIATF, lopezDuterte2020}. To begin analyzing the search trends, there must be two comma separated value (CSV) files per search query: the daily interval time series Google Trends data, $D$, and the weekly data resolution, $W.$ The $D$ file was created from segmenting consecutive 30-day periods that is represented as its own standalone CSV file. Thus, each 30-day period CSV was merged with the rest. Here, the $D_1$ was the initial period commenced on March 16, 2020 until April 15, 2020. Subsequent periods were defined using the same 30-day duration, with the second period, $D_2$, spanned from April 16, 2020 to May 16, 2020. This sequential partitioning shall continue until it has reached $D_k,$ wherein $k$ is the last iteration of the daily resolution data, provided that the last day is March 15, 2021 on that CSV file. There is also $W,$ yet this requires no intervention because this only needs to be retrieved from Google Trends as a year-long weekly data.

The problem on each $D_y$ is that its Google Trends daily RSV cannot be directly compared against another 30-day segment because its respective values are locally relative to each other. For example, a value of 80 in the period of March 16, 2020 until April 15, 2020 is not necessarily equal to an 80 in the period of April 16, 2020 to May 16, 2020. This is due to their time periods being mutually exclusive from one another. In aid of this problem, it must determine the search interest weight, $x_{i, j},$ such that $i$ represents the keyword, and $j$ is the ordinal number of the week from the one-year timeline of March 16, 2020 to March 15, 2021 \cite{brodeurCOVID192021}. The interest weight will be multiplied on the original daily data to have it rescaled to a year-long temporal length, but the RSV values are still showcased on a daily basis.

The first step in Algorithm~\ref{alg:rescaling} is the \texttt{CalculateWeeklyMetrics} function that processes aggregated weekly data $W$ via iterating it per time interval. Then, it proceeds to identify its start and end dates, and computing the sum, count, and average of the daily values within that particular week. The second procedure, \texttt{CalculateWeights}, determines scaling factors for each week by calculating the ratio of the weekly value that was retrieved directly from Google Trends to its average weekly value from the daily data. It defaults to $1$ when the average is zero to avoid undefined numbers. Finally, each daily data point is adjusted in the \texttt{RescaleValues} procedure using the weight of its corresponding week, and preserving the original value when the weekly average is zero. This finalizes on how to circumvent the limitation of a particular user not be able to download daily resolution data from Google Trends beyond nine months.

\subsection{Second Stage: Compute Correlation and Adjacency Matrix}\label{sec:second-corr}

The \texttt{dcor} Python (version 3.12.2) package was imported to apply the distance correlation of each keyword \cite{ramos-carrenoDcor2023}. The result was a 15 by 15 correlation matrix \mycomment{$\mathbf{R}$}wherein each keyword was compared against one another. The distance correlation is within the continuous values of $0$ and $1$. This was chosen against the Pearson correlation coefficient due to the following robust characteristics: it could measure an arbitrary number of dimensions between two random variables (including on the circumstance that both variables are not injective), both comparisons are independent to one another if and only if the correlation is $0,$ and the distance correlation is more effective on comparing on two possible nonlinear relationships \cite{ratnasingamDistance2023}. It is denoted that while the Pearson correlation does not require for the data to have a normal distribution, it is not suggested to be applied on data that are non-normally distributed \cite{houDistance2022}.

\mycomment{\begin{equation}
\mathbf{R} = 
\begin{bmatrix}
1 & r_{k_1,k_2} & r_{k_1,k_3} & \cdots & r_{k_1,k_{15}} \\
r_{k_2,k_1} & 1 & r_{k_2,k_3} & \cdots & r_{k_2,k_{15}} \\
r_{k_3,k_1} & r_{k_3,k_2} & 1 & \cdots & r_{k_3,k_{15}} \\
\vdots & \vdots & \vdots & \ddots & \vdots \\
r_{k_{15},k_1} & r_{k_{15},k_2} & r_{k_{15},k_3} & \cdots & 1
\end{bmatrix}
\end{equation}
\begin{center}
\text{\footnotesize where $r_{k_i,k_j}$ is the distance correlation coefficient between keywords $k_i$ and $k_j$, $i,j \in \{1,\ldots,15\}$}
\end{center}}

The correlation was also administered either on a 15- or 30-day window. This means, the vector space of both keywords were RSV indices from the previous 15 or 30 days on an overlapping and rolling basis. This served the foundation of the correlation score of that window's final day. The adjacency matrix was created thereafter to evaluate on what keywords would form an ``edge'' to create the network graph. The piecewise function fills up the adjacency matrix whether the keyword merits a 0 (no edge) or 1 (with an edge) against its corresponding keyword. This function relies on the thresholds of $\theta=\{0.4, 0.5, 0.6, 0.8\}$. These values for the parameter were considered so that there is an opportunity to investigate on how the network statistics behave besides the standard threshold of 0.5 \cite{soVisualizing2020}.

\mycomment{\begin{equation}
f(x, \theta) = 
\begin{cases}
1, & \text{if } x \geq \theta \\
0, & \text{otherwise}
\end{cases}
\end{equation}}

\subsection{Third Stage: Network Statistic Conversion}\label{sec:third-networkstat}

Two methods of network statistics were used: network density and global clustering coefficient. The network density refers to the proportion of connected edges that currently exist in the graph against the maximum possible number of connections \cite{bedruBig2020}. The network among the search queries is an undirected graph. This means that with $V$ number of nodes (or vertices), there are $\frac{V(V-1)}{2}$ possible edges. Thus, the following is the network density of a graph $D$, such that $E$ represents the existing edges of that network \cite{soTopological2021}:

\begin{equation}
    D = \frac{2E}{V(V-1)}
\end{equation}

On the other hand, the clustering coefficient is a measurement on how probable the nodes of a graph tend to be connected with one another. It pertains to a ratio of the existing number of triangles (three nodes and three edges such that all nodes are connected with one another) over the total number of triplets (three nodes with two edges on that subnetwork). It is defined that $\lambda$ is the number of triangles formed in respect to that singular node (or vertex) $v$ represented as $\lambda(v).$ The number of triples for the entire graph $G$ is $\tau(G) = \sum_{v \in V}\tau(v)$ \cite{chalanconClustering2013}. 

\begin{equation}
    c(G) = \frac{\sum_{v \in V} \lambda(v)}{\tau(G)} = \frac{1}{V}\sum_{v \in V}c(v)
\end{equation}

The rationale behind the network density is to determine how sparse or coagulated the entire keyword network that was constructed. It gives us an idea on how strong the network's predictive power is in terms of the public's Google search behavior at that time. The clustering coefficient in the context of this study is to measure how impactful a keyword is to the density of the entire graph. It generally tells how the keywords used in the network contribute to the graph's predictive behavior power.

\section{Results and Discussion}

\begin{figure}[htbp]
\centering
\includegraphics[width=0.8\textwidth]{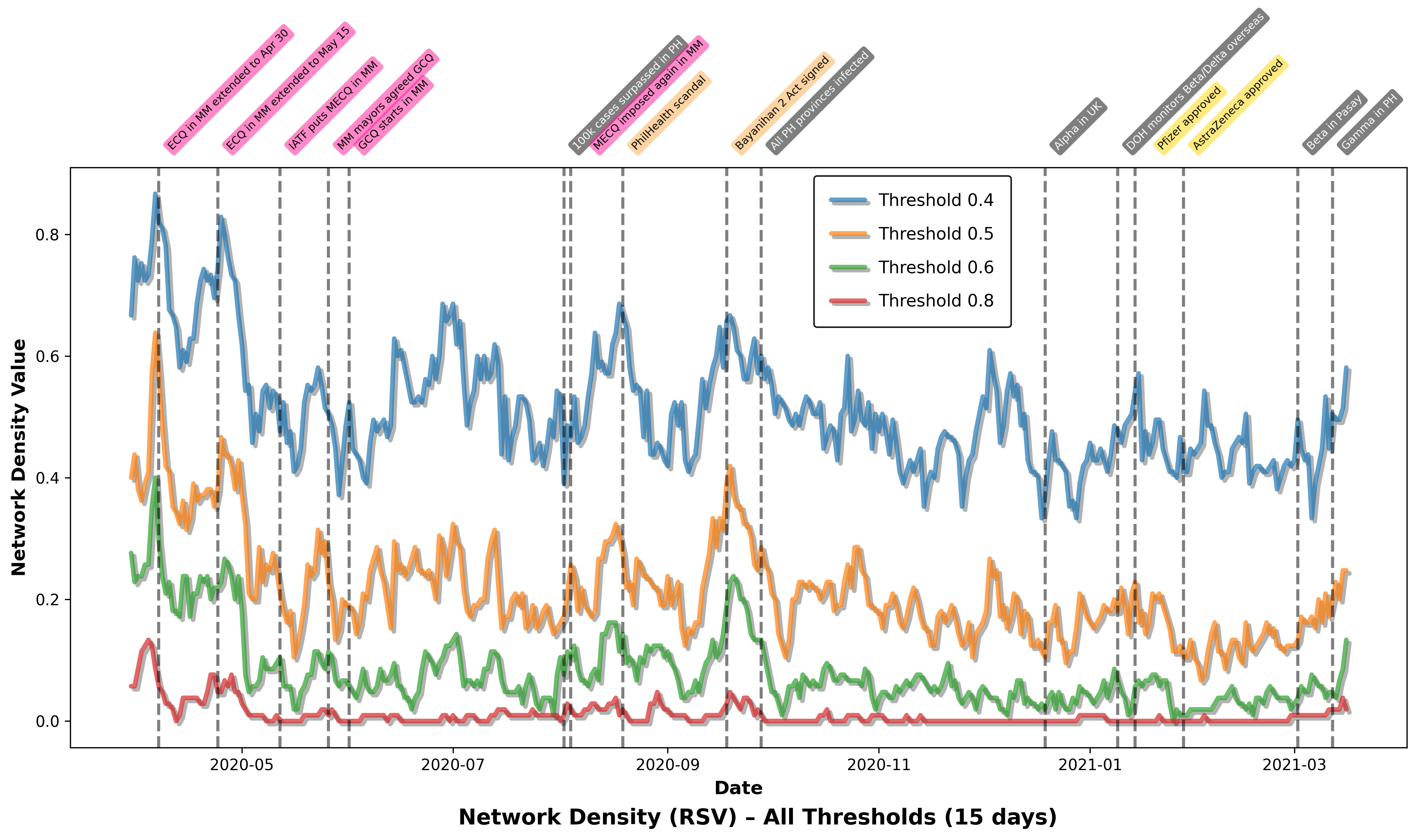}
\vspace{0.3cm}  
\includegraphics[width=0.8\textwidth]{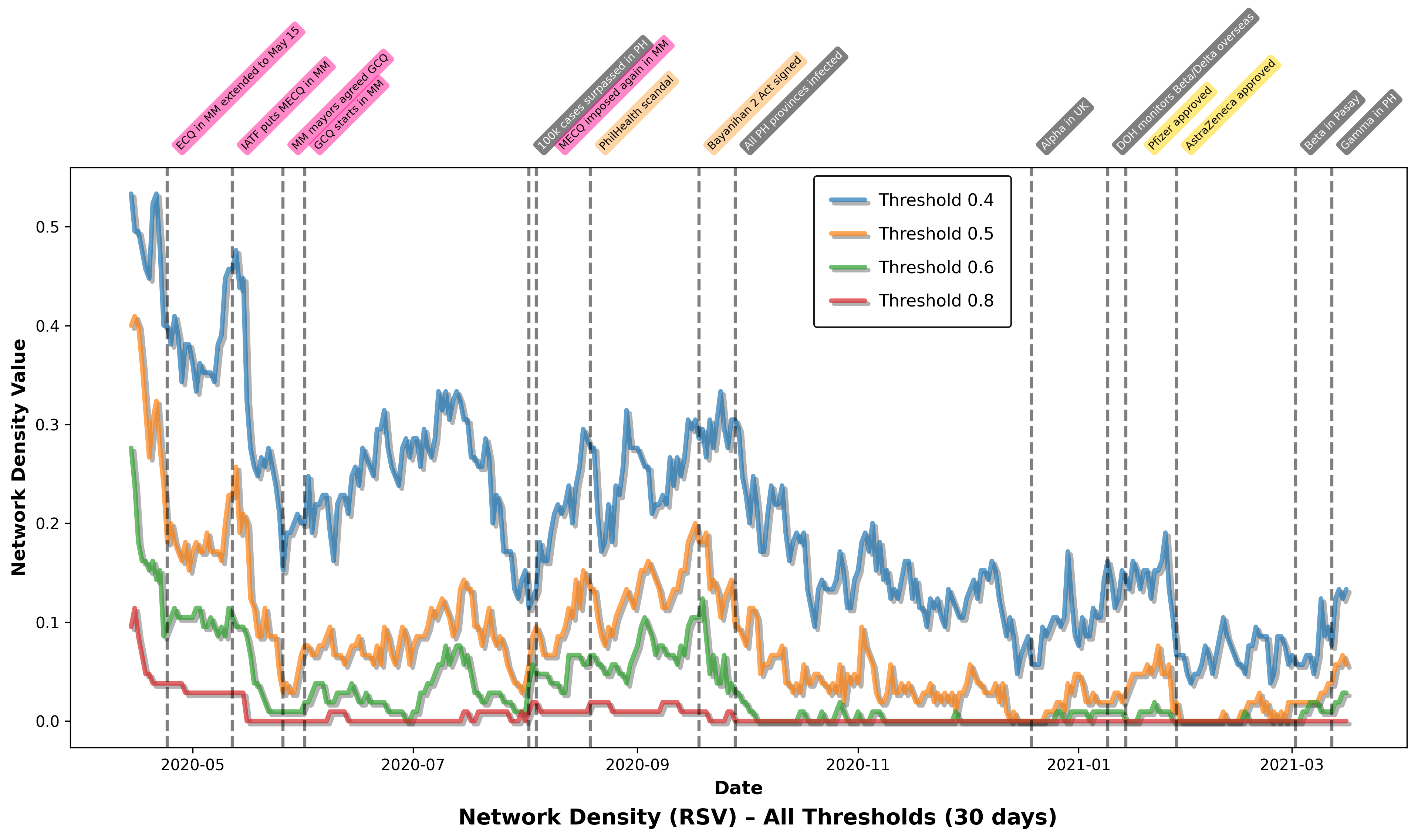}
\caption{Network Density Time-series Line Charts for the 15-Day (top) and 30-Day (bottom) Correlation Window.}
\label{fig:netdense-timeseries}
\end{figure}

\begin{figure}[htbp]
\centering
\includegraphics[width=0.8\textwidth]{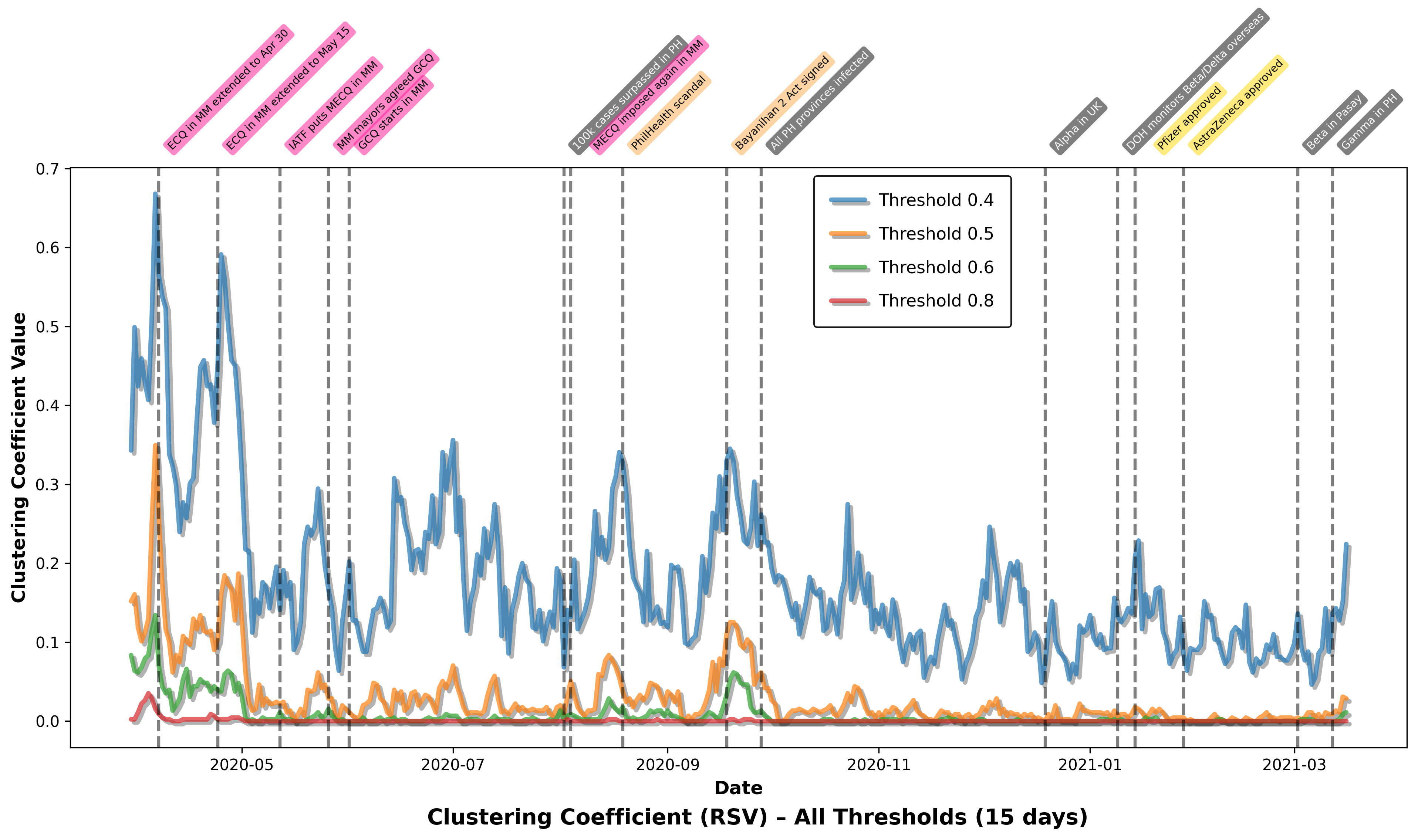}
\vspace{0.3cm}  
\includegraphics[width=0.8\textwidth]{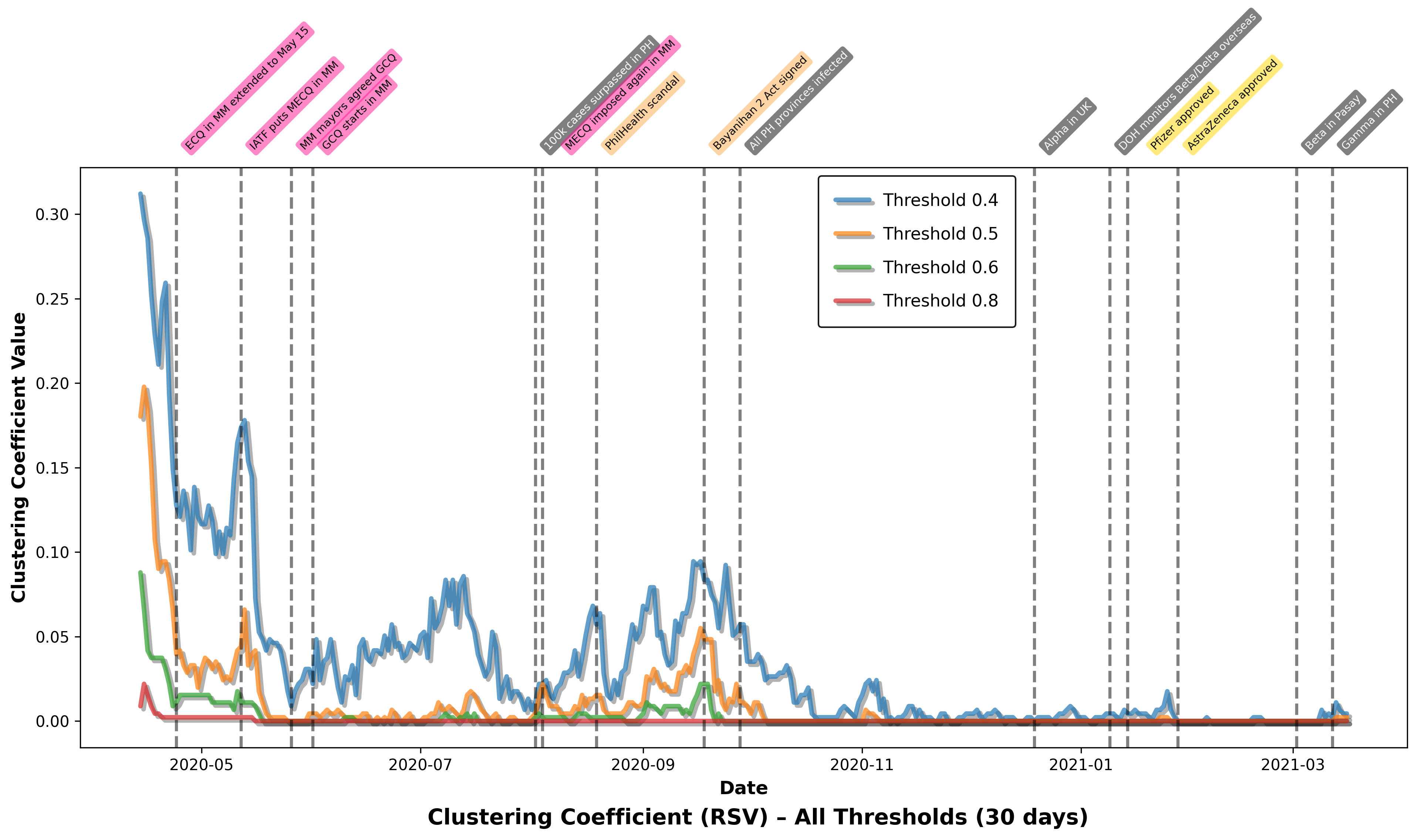}
\caption{Clustering Coefficient Time-series Line Charts for the 15-Day (top) and 30-Day (bottom) Correlation Window.}
\label{fig:cluscoeff-timeseries}
\end{figure}

The four time-series outputs are in terms of whether the correlation was conducted within a 15- or 30-day rolling window, and if it measures the network density or clustering coefficient of that network. The length of the time-series is dependent on the correlation rolling window applied. If it is a 15-day window, the span is on March 31, 2020 until March 16, 2021 because the correlation value for the March 31, 2020 day was dependent on the RSV of the previous 15 days leading up to that point. Logically, the 30-day window lasts on April 15, 2020 until March 16, 2021. Figure \ref{fig:netdense-timeseries} (network density) and Figure \ref{fig:cluscoeff-timeseries} (clustering coefficient) feature the time-series line charts of each network statistic.

The vertical dashed lines correspond to a concerning event that occurred during the pandemic. Its purpose to provide additional details on how the time-series graph was developed in relation with the socio-political context. The color labels for these vertical dashed lines are: magenta pertains to the community quarantines or lockdowns imposed by the public officials or health authorities in the Metro Manila region, black refers to the grim milestones of the pandemic in the Philippines, and the orange and yellow colors represent the political/legislative events, and the approved vaccines in relation to the COVID-19 health crisis respectively. Such events were collected from the reports of reputable news organizations in the Philippines \cite{merezTIMELINE2024, bacligTIMELINE2021}. Table~\ref{tab:events} specifically listed the different events.

\begin{table}[htbp]
\caption{Timeline of Significant Events of National COVID-19 Pandemic Concern.}
\label{tab:events}
\small
\begin{tabular}{lll}
\hline
\textbf{Date} & \textbf{Event} & \textbf{Category} \\
\hline
April 7, 2020 & Enhanced Community Quarantine (ECQ) & Magenta (Quarantine) \\
        & in Metro Manila extended to Apr 30 & \\
April 24, 2020 & ECQ in Metro Manila extended to May 15 & Magenta (Quarantine) \\
May 12, 2020 & IATF puts Modified Enhanced Community & Magenta (Quarantine) \\
             & Quarantine (MECQ) in Metro Manila & \\
May 26, 2020 & Metro Manila elected mayors agreed & Magenta (Quarantine) \\
             & General Community Quarantine (GCQ) & \\
June 1, 2020 & GCQ begins in Metro Manila & Magenta (Quarantine) \\
August 2, 2020 & 100,000 COVID-19 recorded cases & Black (Milestone) \\
             & surpassed nationwide & \\
August 4, 2020 & MECQ imposed again in Metro Manila & Magenta (Quarantine) \\
August 19, 2020 & Philippine Health Insurance Corporation & Orange (Policy) \\
             & (PhilHealth) alleged corruption scandal & \\
September 18, 2020 & Second emergency ``Bayanihan'' National & Orange (Policy) \\
             & Law signed (financial stimulus package) \\
September 28, 2020 & All Philippine provinces infected & Black (Milestone) \\
             & with COVID-19 & \\
December 19, 2020 & Alpha variant detected in United Kingdom & Black (Variant) \\
January 9, 2021 & National government monitors & Black (Variant) \\
             & Beta/Delta variants overseas & \\
January 14, 2021 & Pfizer COVID-19 vaccine approved for & Yellow (Vaccine) \\
             & emergency use & \\
January 28, 2021 & AstraZeneca COVID-19 vaccine approved & Yellow (Vaccine) \\
             & for emergency use & \\
March 2, 2021 & Beta variant detected in Pasay City & Black (Variant) \\
             & (Metro Manila) & \\
March 12, 2021 & Gamma variant detected in the Philippines& Black (Variant) \\
\hline
\end{tabular}
\end{table}

Both network statistics have three generalized observations. First, the threshold parameters have an inverse effect in relation to how dense and clustered the network graph is. To compare in terms of the 15-day correlation window, the peak density value for the 0.4 threshold is 0.8667, for the 0.5 threshold is 0.6381, and for the 0.6 threshold is at 0.4000. These three threshold parameters had their highest density value on April 6, 2020. Similarly, the density value at 0.8 threshold is 0.1333 (April 4, 2020). Such observation is also true for its 30-day correlation window counterpart. \mycomment{0.6762 (0.4 threshold), 0.5429 (0.5 threshold), 0.3238 (0.6 threshold), and 0.0952 (0.8 threshold) wherein all of them occurred on April 14, 2020.} In terms of the clustering coefficient at 15-day correlation window, the highest values (per thresholds) are the following: 0.5714 on March 31, 2020 (0.4 threshold), 0.2945 on April 6, 2020 (0.5 threshold), 1.780 on April 6, 2020 (0.4 threshold), and 0.0219 on April 15, 2020 (0.8 threshold). As expected, this pattern could also be seen for the 30-day window. \mycomment{0.4351 (0.4 threshold), 0.3231 (0.5 threshold), 0.1077 (0.6 threshold), and 0.0109 (0.8 threshold).} The consistent inverse relationship demonstrates across different threshold values suggests that stronger correlations between search terms were increasingly rare, but potentially becomes more meaningful. For example, at the 15-day correlation window with a 0.8 threshold, the most common connections were: ``fever-flu'' and ``cough-flu'' (29 connections each) from April to June 2020, ``quarantine-ecq'' (52 connections) from July to September 2020, ``frontliners-masks'' (4 connections) from October to December 2020, and ``ecq-lagnat'' (16 connections) from January to March 2021. These keyword partnerships show the core search behavior because all of them persisted under the most stringent threshold requirements. In other words, it is \emph{likelier} that these words were looked up by those residing in Metro Manila during its respective time period.

Relating with the first observation, another noticeable trend concerns an initial high network and clustering value coefficients during the pandemic's first months. During early 2020, the 15-day RSV networks exhibited exceptionally high network density, reaching a peak of 0.8667 on April 6. This was paralleled by a high clustering coefficient of 0.6681 on the same date. These concurrent peaks in both metrics indicate not only a high degree of overall connectivity between search terms (density) but also a strong tendency for these terms to form tightly interconnected local cliques (clustering coefficient). Such alignment suggests that users were not only searching for many related terms but were also exploring these terms in highly structured patterns. The high network density indicates that many search terms were being used together frequently, while the high clustering coefficient suggests these searches were occurring in well-defined thematic groups. For instance, the sustained high density values throughout early April 2020 (e.g., 0.8190 on April 7, and 0.8095 on April 8) coupled with strong clustering coefficients (0.5648 and 0.5385 respectively) demonstrate how search behaviors were both extensive and highly organized during this critical period.

Both metrics showed a coordinated decline when the pandemic progressed, though following slightly different patterns. The network density experienced a more gradual decrease, maintaining relatively high values (above 0.6) through April, while clustering coefficients showed a more pronounced decline. This difference in decay rates suggests that while overall search term relationships remained relatively strong, the tight local clustering of terms began to disperse more quickly. By May 2020, both metrics had significantly decreased, with network density falling below 0.5 and clustering coefficients dropping to around 0.3 to 0.4, indicating a transition to more diverse and less structured search patterns. The temporal evolution of both network metrics in the 30-day windows provides additional insight into the long-term dynamics of information-seeking behavior. The 30-day RSV networks showed lower peak values for both density (0.5333 on April 14) and clustering coefficient (0.3121 on April 14), but maintained more stable values over time. This stability in longer time windows suggests that while immediate search behaviors were highly focused and clustered, the underlying pattern of information seeking evolved more gradually.

Lastly, there were distinct temporal patterns in the health-seeking behavior, characterized by evolving relationships between health-related and policy-related queries. In April to June 2020, the network exhibited its strongest interconnections, with the highest number of consistent keyword pairs and triads. At the 0.4 threshold, the strongest keyword pair was ``quarantine-ecq'' with 85 connections, followed by ``social distancing-work from home'' with 82 connections, indicating the people's focus on policy measures and lifestyle adjustments. The triad analysis during this period revealed strong interconnections between healthcare terms and policy measures, with ``frontliners-social distancing-work from home'' appearing 60 times. As the pandemic progressed into the middle period (July-September 2020), the network structure showed a shift in focus, with ``quarantine-ecq'' maintaining prominence (92 connections) but new combinations emerging, particularly involving protective equipment terms. The introduction of ``face shield'' into frequent pairs and triads (e.g., ``quarantine-face shield-ecq'') reflects the evolution of the protocols. This period also showed strong sustained connections between symptom-related terms (fever, flu, cough). The later periods (October-December 2020 and January-March 2021) demonstrated a gradual transformation in search patterns. The network showed increased fragmentation at higher thresholds (0.6 and 0.8). Notable was the emergence of stronger connections between Filipino terms for symptoms (sipon, ubo, lagnat) and policy measures. 

By February 2021, the network structure had significantly evolved, with pairs like ``headache-sipon'' (70 connections) and ``frontliners-fever'' (60 connections) dominating, and witnessed a shift from policy-focused searches to more symptom-specific queries. The analysis across different thresholds (0.4 to 0.8) reveals the hierarchy of search term relationships. While the 0.4 threshold captured broader associative patterns, the persistence of certain connections at higher thresholds (particularly symptom-related terms) indicates fundamental relationships in public health information seeking. The 30-day window analysis showed more stable but less intense connections compared to the 15-day window, suggesting that immediate search behaviors were more strongly correlated than longer-term patterns.

\section{Conclusion and Future Work}

This study revealed significant patterns in public search behavior during the COVID-19 pandemic via the analysis of a network density perspective of Google Trends data. Three key findings were discovered. It was first observed that there was a clear indirect relationship between threshold values and both network density and clustering coefficients. As the correlation threshold increased from 0.4 to 0.8, network metrics showed systematic decreases across all time periods. This phenomenon is an evidence that while weaker correlations were common, stronger relationships between search terms were rarer yet potentially more meaningful. Second, the temporal analysis revealed exceptionally high network connectivity during the initial months of the pandemic. This period of intense network cohesion was characterized by strongly connected search patterns. It suggests a highly focused and synchronized public attention on pandemic-related information. Yet as time passed, the gradual decline in both metrics indicates an evolution from concentrated crisis-response information seeking to more diverse Google searching. Lastly, the analysis of keyword connections and triadic relationships revealed emerging patterns of public interest. Some examples of those patterns were the protective equipment terms and symptom-specific searches. This is evidence that it maintained strong connections between core health-related terms. The later periods however showed increased fragmentation and diversification, especially at higher thresholds.

These results have important implications for public health communication and crisis response. The strong initial network connectivity suggests a critical window for effective information dissemination during early health crisis periods. This is also evidence of expressed uncertainty among communities due to their predictive search activity in the first few months of the pandemic. Then, the persistence of certain keyword relationships across thresholds indicates fundamental topics that remain relevant throughout a health crisis. Lastly, the evolution of search patterns over time suggests the need for more strategic communication plans towards mitigating the cases that adapt to the changing public information needs. Future considerations include to investigate the forecasting capabilities of network metrics in terms of future disease cases, and examine the relationship between search network structure and public health outcomes. Overall, this work establishes one of the foundations for understanding collective health-seeking behavior during public health crises from a network perspective.

\section*{Acknowledgements}

The authors express its gratitude to the Ateneo Social Computing Science Laboratory, with its parent entity, the Ateneo Center for Computing Competency and Research (ACCCRe) for the support, opportunity, and avenue to fulfill this study. Much appreciation is also extended to Christian E. Pulmano and Kennedy E. Espina for the intellectual wisdom imparted.

\bibliography{references}

\begin{thebibliography}{10}

\bibitem{amitEarly2021}
Amit AM, Pepito VC, Dayrit M.
\newblock Early Response to {{COVID-19}} in the {{Philippines}}.
\newblock West Pac Surveill Response J. 2021 Mar;12(1):56-60.

\bibitem{bbcnewsCoronavirus2020}
{BBC News}. Coronavirus: {{Millions}} Return to Lockdown in {{Philippines}}; 2020.
\newblock https://www.bbc.com/news/world-asia-53646149.

\bibitem{delcastilloEstimating2024}
Del~Castillo MFP, Fujimi T, Tatano H.
\newblock Estimating Sectoral {{COVID-19}} Economic Losses in the {{Philippines}} Using Nighttime Light and Electricity Consumption Data.
\newblock Front Public Health. 2024 Feb;12.

\bibitem{alibudbudCase2023}
Alibudbud R.
\newblock A {{Case}} of {{Pharmaceutical Messianism Amidst}} the {{COVID-19 Pandemic}}: {{An Infodemiological Study}} of {{Ivermectin}} in the {{Philippines}}.
\newblock Policy Politics Nurs Pract. 2023 Feb;24(1):17-25.

\bibitem{mallariMediating2023}
Mallari EFI, Ato CKA, Crucero LJMO, Escueta JT, Eslabra VAP, Urbano PEM.
\newblock The Mediating Role of Impulse Buying on Hedonic Shopping Motivation and Life Satisfaction of Online Shoppers in the {{Philippines}}.
\newblock Int Soc Sci J. 2023 Sep;73(249):861-72.

\bibitem{alibudbudGoogle2023}
Alibudbud R.
\newblock Google {{Trends}} for Health Research: {{Its}} Advantages, Application, Methodological Considerations, and Limitations in Psychiatric and Mental Health Infodemiology.
\newblock Front Big Data. 2023;6.

\bibitem{rovettaGoogle2024}
Rovetta A.
\newblock Google Trends in Infodemiology: {{Methodological}} Steps to Avoid Irreproducible Results and Invalid Conclusions.
\newblock Int J Med Inform. 2024 Oct;190.

\bibitem{cebrianAddressing2024}
Cebri{\'a}n E, Domenech J.
\newblock Addressing {{Google Trends}} Inconsistencies.
\newblock Technol Forecast Soc Change. 2024 May;202.

\bibitem{lolicDIY2024}
Loli{\'c} I, Mato{\v s}ec M, Sori{\'c} P.
\newblock {{DIY}} Google Trends Indicators in Social Sciences: {{A}} Methodological Note.
\newblock Technol Soc. 2024 Jun;77.

\bibitem{behnenExperimental2020}
Behnen P, Kessler R, Kruse F, G{\'o}mez JM, Schoenmakers J, Zerr S.
\newblock Experimental {{Evaluation}} of {{Scale}}, and {{Patterns}} of {{Systematic Inconsistencies}} in {{Google Trends Data}}.
\newblock In: Koprinska I, Kamp M, Appice A, Loglisci C, Antonie L, Zimmermann A, et~al., editors. Workshops of the {{European Conference}} on {{Machine Learning}} and {{Knowledge Discovery}} in {{Databases}} ({{ECML PKDD}} 2020). vol. 1323 of Communications in {{Computer}} and {{Information Science}}. Ghent, Belgium: Springer International Publishing; 2020. p. 374-84.

\bibitem{mavraganiGoogle2019}
Mavragani A, Ochoa G.
\newblock Google {{Trends}} in {{Infodemiology}} and {{Infoveillance}}: {{Methodology Framework}}.
\newblock JMIR Public Health Surveill. 2019 May;5(2).

\bibitem{eichenauerObtaining2022}
Eichenauer VZ, Indergand R, Mart{\'i}nez IZ, Sax C.
\newblock Obtaining Consistent Time Series from {{Google Trends}}.
\newblock Econ Inq. 2022;60(2):694-705.

\bibitem{medeirosProper2021}
Medeiros MC, Pires HF. The {{Proper Use}} of {{Google Trends}} in {{Forecasting Models}}. arXiv; 2021.

\bibitem{chuEnhancing2023}
Chu AMY, Chong ACY, Lai NHT, Tiwari A, So MKP.
\newblock Enhancing the {{Predictive Power}} of {{Google Trends Data Through Network Analysis}}: {{Infodemiology Study}} of {{COVID-19}}.
\newblock JMIR Public Health Surveill. 2023 Sep;9.

\bibitem{heveyNetwork2018}
Hevey D.
\newblock Network Analysis: A Brief Overview and Tutorial.
\newblock Health Psychol Behav Med. 2018 Jan;6(1):301-28.

\bibitem{bedruBig2020}
Bedru HD, Yu S, Xiao X, Zhang D, Wan L, Guo H, et~al.
\newblock Big Networks: {{A}} Survey.
\newblock Comput Sci Rev. 2020 Aug;37.

\bibitem{soTopological2021}
So MKP, Chu AMY, Tiwari A, Chan JNL.
\newblock On Topological Properties of {{COVID-19}}: Predicting and Assessing Pandemic Risk with Network Statistics.
\newblock Sci Rep. 2021 Mar;11(1).

\bibitem{departmentofhealthoftherepublicofthephilippinesRecommendations2020}
{Department of Health of the Republic of the Philippines}. Recommendations for the {{Management}} of the {{Coronavirus Disease}} 2019 ({{COVID-19}}) {{Situation}}; 2020.

\bibitem{Pytrends}
Pytrends: {{Pseudo API}} for {{Google Trends}};.
\newblock Available from: \url{https://pypi.org/project/pytrends/}.

\bibitem{philippinestatisticsauthorityMore2023}
{Philippine Statistics Authority}. More than 50 Million Have {{Access}} to the {{Internet}} (2020 {{Census}} of {{Population}} and {{Housing}}); 2023.

\bibitem{worldhealthorganizationAdvice2023}
{World Health Organization}. Advice for the Public on {{COVID-19}} -- {{World Health Organization}}; 2023.
\newblock https://www.who.int/emergencies/diseases/novel-coronavirus-2019/advice-for-public.

\bibitem{worldhealthorganizationCoronavirus2023}
{World Health Organization}. Coronavirus Disease ({{COVID-19}}); 2023.
\newblock https://www.who.int/news-room/fact-sheets/detail/coronavirus-disease-(covid-19).

\bibitem{departmentofhealthoftherepublicofthephilippinesIATF}
{Department of Health of the Republic of the Philippines}. {{IATF Resolution Archives}} ({{March}} to {{April}} 2020);.
\newblock https://doh.gov.ph/iatf/page/32.

\bibitem{lopezDuterte2020}
Lopez V. Duterte Convenes Inter-Agency Body as {{COVID-19}} Cases Rise to 20; 2020.
\newblock https://www.gmanetwork.com/news/topstories/nation/728975/duterte-convenes-inter-agency-body-as-covid-19-cases-rise-to-20/story/.

\bibitem{brodeurCOVID192021}
Brodeur A, Clark AE, Fleche S, Powdthavee N.
\newblock {{COVID-19}}, Lockdowns and Well-Being: {{Evidence}} from {{Google Trends}}.
\newblock J Public Econ. 2021 Jan;193.

\bibitem{ramos-carrenoDcor2023}
{Ramos-Carre{\~n}o} C, Torrecilla JL.
\newblock Dcor: {{Distance}} Correlation and Energy Statistics in {{Python}}.
\newblock SoftwareX. 2023 May;22.

\bibitem{ratnasingamDistance2023}
Ratnasingam S, {Mu{\~n}oz-Lopez} J.
\newblock Distance {{Correlation-Based Feature Selection}} in {{Random Forest}}.
\newblock Entropy. 2023 Sep;25(9).

\bibitem{houDistance2022}
Hou J, Ye X, Feng W, Zhang Q, Han Y, Liu Y, et~al.
\newblock Distance Correlation Application to Gene Co-Expression Network Analysis.
\newblock BMC Bioinform. 2022 Dec;23(1).

\bibitem{soVisualizing2020}
So MKP, Tiwari A, Chu AMY, Tsang JTY, Chan JNL.
\newblock Visualizing {{COVID-19}} Pandemic Risk through Network Connectedness.
\newblock Int J Infect. 2020 Jul;96:558-61.

\bibitem{chalanconClustering2013}
Chalancon G, Kruse K, Babu MM.
\newblock Clustering {{Coefficient}}.
\newblock In: Dubitzky W, Wolkenhauer O, Cho KH, Yokota H, editors. Encyclopedia of {{Systems Biology}}. New York, NY: Springer; 2013. p. 422-4.

\bibitem{merezTIMELINE2024}
Merez A, Sabillo K, Ramos P. {{TIMELINE}}: {{The}} Coronavirus Disease Crisis in the {{Philippines}}; 2024.

\bibitem{bacligTIMELINE2021}
Baclig CE. {{TIMELINE}}: {{One}} Year of {{Covid-19}} in the {{Philippines}}; 2021.
\newblock https://newsinfo.inquirer.net/1406004/timeline-one-year-of-covid-19-in-the-philippines.

\end{thebibliography}

\end{document}